\documentclass{CEAS_EuroGNC_v1.0}

\usepackage[utf8]{inputenc}

\usepackage{graphicx}
\usepackage{amsmath}
\usepackage[version=4]{mhchem}
\usepackage{siunitx}
\usepackage{array}
\usepackage{longtable}
\usepackage{bm}
\usepackage[short]{optidef}
\usepackage{comment}

\newcommand{\rs}[1]{_\mathrm{#1}}

\newcommand{\summa}[2]{\sum_{#1}^{#2}}
\newcommand{\tr}{^\top}

\title{Reconfiguration of a satellite constellation in circular formation orbit with decentralized model predictive control}

\ifpdf
\hypersetup{pdfinfo={
Title={Reconfiguration of a satellite constellation in circular formation orbit with decentralized model predictive control},
Author={Tomas Pippia, Valentin Preda, Samir Bennani, and Tamas Keviczky},
Subject={Reconfiguration of a satellite constellation in circular formation orbit with decentralized model predictive control},
Keywords={Model Predictive Control; Satellite constellation; Decentralized control}
}}
\fi

\begin{document}
\maketitle

\begin{authorList}{4cm}
\addAuthor{Tomas Pippia}{Postdoctoral researcher, Delft Center for Systems and Control, Delft University of Technology, 2600AA, Delft, The Netherlands. \emailAddress{tomas.pippia@gmail.com}}
\addAuthor{Valentin Preda}{GNC System Engineer, European Space Agency (ESA), European Space Research and Technology Centre (ESTEC), 2201 AZ, Noordwijk, The Netherlands \emailAddress{valentin.preda@esa.int}}
\addAuthor{Samir Bennani}{Head of the GNC, AOCS and Pointing Division (ESTEC), European Space Agency (ESA), European Space Research and Technology Centre (ESTEC), 2201 AZ, Noordwijk, The Netherlands
\emailAddress{samir.bennani@esa.int}}
\addAuthor{Tamas Keviczky}{Full Professor, Delft Center for Systems and Control, Delft University of Technology, 2600AA, Delft, The Netherlands. \emailAddress{t.keviczky@tudelft.nl}}
\end{authorList}

\begin{abstract} 
Satellite constellation missions, consisting of a large number of spacecraft, are increasingly being launched or planned. Such missions require novel control approaches, in particular for what concerns orbital phasing maneuvers. In this context, we consider the problem of reconfiguration of a satellite constellation in a circular formation. In our scenario, a formation of equally spaced spacecraft need to undergo an autonomous reconfiguration due to the deorbiting of a satellite in the formation. The remaining spacecraft have to reconfigure to form again an equidistant formation.
To achieve this goal, we consider two decentralized strategies that rely on different sets of information about the neighboring spacecraft in the formation. In the fully decentralized case, each controller knows only the current states of each spacecraft, i.e.\ position and velocity, while in the second decentralized strategy with with information sharing, the entire planned nominal trajectory of each spacecraft is available to its neighbors.
Our numerical simulation results show that, by increasing the amount of information available to each spacecraft, faster reconfiguration maneuvers with smaller fuel consumption can be achieved.
\end{abstract}

\keywords{Model Predictive Control; Satellite constellations; Spacecraft guidance, navigation, and control}



\section{Introduction}\label{sec:introduction}

In recent years, there has been an increase in the number of missions and studies regarding satellite constellations, performed by both industry \cite{starlink2020,iridium2021} and academia \cite{vandam2019b,hadaegh2016,diMauro2018,bandyopadhyay2016b,sinYin2020,vos2014, morgan2014a,morgan2016,foust2020a}. Such systems need to achieve a mission goal, which can vary from commercial applications, e.g., internet-from-space, to scientific missions, such as in-orbit interferometers. Given the large amount of spacecraft in such missions and the goal to be achieved, autonomous coordination between the satellites is needed. 
Indeed, while for single monolithic satellites or for a handful of spacecraft a centralized control system would be applicable and safe, the same kind of control strategy applied to large constellations would be impractical and most likely infeasible \cite{chen2018,foust2020a,morgan2014a,morgan2016}. By having a single central node that computes the commands for the whole constellation, the system becomes less robust to communication issues, such as delays, and to failures, as a fault in the central node compromises the whole mission. Moreover, if the satellites are controlled by a ground station, the amount of possible collision avoidance maneuvers to be performed will increase exponentially with the amount of spacecraft. In fact, recently an ESA satellite was forced to perform a maneuver to avoid a Starlink constellation satellite \cite{esaavoidance}. Lastly, the ground station might be missing if the constellation is deployed around another celestial object, e.g.\, Mars \cite{sinYin2020}. Therefore, an autonomous controller is needed for the new generation of satellite constellations.

Thanks to non-centralized control strategies, many new types of missions are possible, e.g., in-orbit interferometers or distributed sensors \cite{hadaegh2016,montenbruck2011}. However, in order to be successful, these constellations require adequate controllers that can achieve stable and safe performance.
In this regard, decentralized or distributed control schemes can mitigate most of the aforementioned issues \cite{hadaegh2016,negenborn2014}. The key idea is to have a controller on-board each spacecraft, so that the control action is computed locally. In this way, the computational complexity related to the computation of the control action would be much lower \cite{morgan2012b}. Moreover, while communication issues, e.g., delays, or a sudden disconnection, would still be present, they would have a much reduced impact. Indeed, centralized controllers rely on the availability of information from the whole constellation and the lack of communication from few faulty nodes could quickly lead to instability or infeasibility. On the other hand, in non-centralized controllers, the spacecraft only communicate with the other neighboring satellites, thus a faulty communication link has only a local impact. Robustness of the system would be increased as well: even in the case where some satellites malfunction, the other ones would still be able to compute a proper control action \cite{diMauro2018,koenig2018,hadaegh2016}. Furthermore, a single faulty satellite could be easily replaced by another one, so that the whole mission is not compromised, whereas a failure in the central node of a centralized controller might have catastrophic results for the mission. 

Among the several non-centralized controllers, Model Predictive Control (MPC) stands out as one of the main control tools for spacecraft \cite{camacho2013}. MPC has been successfully applied to constellation of spacecraft in previous works \cite{morgan2014a,morgan2016,foust2020a}. In \cite{morgan2014a}, authors present a decentralized MPC method to perform a reconfiguration of spacecraft. The authors rely on a linearized spacecraft motion model at every time step and convexify collision avoidance constraints, employing a sequential convex programming approach. At each time step, the spacecraft in the constellation receive the nominal trajectories from their neighbors. The collision avoidance constraints are then updated and cast in a convex form, so that a convex problem is solved online. In \cite{morgan2016}, the algorithm discussed in \cite{morgan2014a} is applied together with an online assignment of the final desired position of each spacecraft in the formation. Such positions are obtained via an online auction scheme. The authors of \cite{foust2020a} also use the sequential convex programming of \cite{morgan2014a} but they augment it with a method for in-orbit assembly of heterogenous spacecraft, adding therefore the docking capabilities to the formation.

While many articles in the literature have considered an MPC application to a constellation of satellites, most of them study only a ``swarm'' formation, where the satellites fly close to each other. Little attention has been given to satellites in a circular orbit formation, which is not only of academic but also of practical interest. Such formation is indeed applied in internet-from-space missions, e.g.\ \cite{starlink2020}. Furthermore, a circular formation could also be applied in other planets to have ground coverage at all times \cite{sinYin2020}. To the best of the authors' knowledge, only \cite{sinYin2020,vos2014} consider such a formation. However, both methods lack some autonomy as they assume that the number of spacecraft is known to every satellite in the formation.

In this paper, we study the application of decentralized MPC to a constellation of satellites equally spaced in a circular formation along a nominal orbit. At one point during mission, some of the spacecraft leave the orbit, thus a gap in the formation is created. The remaining spacecraft have to reconfigure such that a circular formation is kept while also maintaining equal distance from each other.
In our considered scenario, none of the spacecraft knows the total amount of satellites in the constellation.
Such choice makes the problem more challenging, since each spacecraft can only communicate with its closest neighbors or sense them to adjust their position in the orbit. In addition, to verify the importance of intersatellite communication, we analyze both the case in which the spacecraft can communicate with each other and exchange information, i.e., their current \textit{and predicted} position and velocity, and the case in which instead there is no communication and spacecraft can only ``sense'' each other, thus knowing only their current velocity and position.

The outline of the paper is as follows. In Section \ref{sec:model} the models used in this work are explained. Section \ref{sec:control} is devoted to the control design. We present the simulation results in Section \ref{sec:simulations} and lastly conclusions are drawn and the next research steps are presented in Section \ref{sec:conclusions}.

\section{Preliminaries} \label{sec:model}
\subsection{Frames}
We represent the movement of the spacecraft around Earth using two frames, i.e., the Earth-Centered-Inertial (ECI) frame and the Local-Vertical Local-Horizontal (LVLH) frame. The origin of the ECI frame is located at the center of the Earth and the $\hat{X}$-axis points towards the direction of the vernal equinox, the $\hat{Z}$-axis points towards the North pole, and the $\hat{Y}$-axis completes the right-handed coordinate system. The ECI frame is used to locate the (virtual) chief orbit. On the other hand, the LVLH frame is local to every satellite and it is centered at the (virtual) chief spacecraft or orbit. The $\hat{x}$-axis is aligned with the center of the Earth and points away from it, the $\hat{y}$-axis is tangential to the  flight direction along the orbit, and the the $\hat{z}$-axis is aligned with the orbital angular momentum vector of the spacecraft and completes the right-handed coordinate system. 

\subsection{Dynamics}
In order to deal with the large angular distances between spacecraft in the circular formation, we employ the HCW equations in cylindrical coordinates to describe their motion \cite{debruijn2011,geller2017}. The three elements that compose a cylindrical coordinate system are defined w.r.t.\ a reference orbit and are the relative radius $\rho$, the relative angular displacement $\theta$, and the relative elevation $z$. 

Similarly to what is done for the rectilinear HCW equations, we can express the states as relative distances with respect to the (virtual) leader spacecraft. The HCW equations in a cylindrical coordinate system are given by \cite{geller2017}:

\begin{equation}\label{eq:HCWCylindrical}
    \begin{aligned}
    \ddot \rho - 3n^2 \rho - 2Rn \dot \theta & = a_{F_\rho} \\
    \ddot \theta + \frac{2n}{R} \dot \rho & = a_{F_\theta} \\
    \ddot z + n^2 z & = a_{F_z}
    \end{aligned}
\end{equation}
where $R$ is the radius of the (virtual) chief's orbit measured from the center of the Earth, $n$ is the mean motion, and $a_{F_\rho}$, $a_{F_\theta}$, $a_{F_z}$ are the accelerations applied to the spacecraft in the respective directions, including the control input. These equations hold for $\rho \ll R$, $z \ll R$, and $\dot{\theta}\ll n$.

The dynamics \eqref{eq:HCWCylindrical} are very similar to their rectilinear counterpart. There is, however, one fundamental difference. The second-order nonlinear equations of motion, before deriving \eqref{eq:HCWCylindrical} through a first-order Taylor expansion, depend only $\rho$, $\dot{\rho}$, $\dot{\theta}$, $z$. This implies that the linearized equations \eqref{eq:HCWCylindrical} hold for an arbitrarily large $\theta$ and $\dot{z}$ \cite{geller2017}. In other words, the model is valid for large angular displacements between spacecraft, thus making the model very suitable for our circular formation. Moreover, the curvilinear HCW equations achieve a higher orbit prediction accuracy w.r.t.\ their rectilinear counterpart \cite{debruijn2011,hartley2015}.

The dynamics of a single spacecraft $i$ is expressed following \eqref{eq:HCWCylindrical} and by defining a state vector $x_i$ with six components, i.e., $x_i = \begin{bmatrix} \rho_i & \theta_i & z_i & \dot{\rho}_i & \dot{\theta}_i & \dot{z}_i \end{bmatrix}^\top$. Lastly, the continuous-time dynamics are discretized using a simple zero-order hold method, as in \cite{debruijn2011} and can be expressed in matrix form as

\begin{equation}\label{eq:dynamicsLTIHCWRing}
x_i(k+1) = Ax_i + Bu_i(k),
\end{equation}
where $u_i$ are the inputs of spacecraft $i$ and $A$, $B$ are matrices obtainable from \eqref{eq:HCWCylindrical}.

\section{Control algorithm} \label{sec:control}
\subsection{Model predictive control}
In this article, we apply a standard MPC scheme. This means that a finite-time optimal control problem is solved at the current time step over the prediction horizon of the next $N\rs{p}$ steps. The optimization problem yields an optimal control input vector $\bm{u}^*$; we apply only the input corresponding to the current time step. Then, at the next sampling instant, we take new state measurements and solve the updated optimization problem.

\subsection{Constraints}
The only constraint present in our problem, apart from the dynamics \eqref{eq:dynamicsLTIHCWRing}, is the one related to the maximum thrust allowed, defined as 
\begin{equation}\label{eq:umaxRing}
\|\bm{u}(k)\|_\infty \leq u\rs{max}, \ \forall k\in\{1,\ldots,N\rs{p}\}
\end{equation}

In this specific scenario, the distances among spacecraft are very large and thus the satellites do not come close to each other. Therefore, collision avoidance constraints are currently not needed. Even if they were included in the problem, they would make the problem unnecessarily complicated, since they would always be inactive. However, in case such constraints are needed for a specific mission, a scheme as the one in \cite{morgan2014a} can be implemented.

Upper and lower bounds on the state vector $x$ are not included, but they can be added if needed, e.g., if it is necessary to limit the value of $\rho$ to avoid a possible collision with other spacecraft in nearby orbits.

\subsection{Objectives}
As mentioned in Section \ref{sec:introduction}, the objective is to keep the satellites close to the reference orbit and at the same time to maintain an equal distance from each other. Since in the considered scenario we can safely assume that the motion of the spacecraft occurs only in the orbital plane, we can disregard the out-of-plane motion and focus only on the relative radius and relative angle. Note that the state $\rho_i$ tracks the deviation from the reference radius and is therefore related to the orbit-keeping objective. On the other hand, the relative angle state $\theta_i$ is related to the relative displacement between spacecraft and therefore to the equidistance objective.

We can thus use $\rho_i$ as the error of the current orbital radius w.r.t.\ the desired reference one. Therefore, we can use it as an objective function in a quadratic form,
\begin{equation}\label{eq:JrRing}
J^\rho_i(k) = \|\rho_i(k)\|^2_2
\end{equation}
For what concerns the relative angle, we minimize the angular distance of each spacecraft w.r.t.\ the mean angle of its neighbors, contained in the set of neighbors $\mathcal{N}_i$, defined as the set of the $p$-closest neighbors to satellite $i$. For $p = 2$, this means $\mathcal{N}_i$ consists of the satellite ahead and the one behind satellite $i$. We consider therefore the objective
\begin{equation}\label{eq:Jtheta}
J^\theta_i(k) = \|\theta_i(k)-\bar{\theta}_i(k)\|^2_2 
\end{equation}
where $\bar{\theta}_i(k)$ is the setpoint angle and it is computed as
\begin{equation}\label{eq:thetaSetpoint}
\bar{\theta}_i(k) = \frac{1}{|\mathcal{N}_i|} \summa{j\in \mathcal{N}_i}{}\theta_j(k).
\end{equation}

 We would also like to keep the fuel consumption as low as possible. We therefore also consider the cost term
\begin{equation}\label{eq:JuRing}
J^u_i(k) = \|u_i(k)\|^2_2 
\end{equation}

Note that if we use the cost terms \eqref{eq:JrRing}--\eqref{eq:Jtheta} for all the time steps $k+l$, $l \in \{0,\dots,N\rs{p}\}$, we assign the same importance to both the transient and the final position. However, in order to reach their angular setpoint, the spacecraft have to deviate from their radius setpoint by temporarily either increasing or decreasing their orbital radius, i.e., by performing an orbital phasing maneuver. At the same time, the angular setpoint deviation is not important during the transient, but it is our main goal at the end of the horizon.
Therefore, penalizing (too much) the transient could potentially lead to oscillations and waste of fuel because the spacecraft would try to stay as close as possible to their setpoint \textit{at all times}, rather than assigning more importance to the ultimate goal of reaching equidistance. On the other hand, not penalizing the transient at all could be detrimental, since we would allow the spacecraft to deviate too much from their setpoints before the final time step. In order to prevent these issues, we penalize the final position with a weight $\alpha\rs{end}$, which can be tuned to achieve the desired performance. Furthermore, during the transient we do not penalize the angular setpoint, as our focus and main goal is the final relative angular position. We still penalize the radius setpoint during the transient to avoid that the spacecraft deviate too much from their reference orbit.

We therefore obtain a cost function for the transient as a total weighted cost
\begin{equation}\label{eq:j_transient}
J_i^{\mathrm{trns}}(k) = \alpha_\rho J^\rho_i(k) + \alpha_u J^u_i(k),
\end{equation}
and a cost function for the final position, i.e.\
\begin{equation}\label{eq:j_final_position}
J_i^{\mathrm{end}}(k) = \alpha_\theta J^\theta_i(k) + \alpha_\rho J^\rho_i(k) ,
\end{equation}
where $\alpha_\theta>0$, $\alpha_\rho>0$, and $\alpha_u>0$ are weights that can be tuned to achieve the desirable performance.

\subsection{Optimization problem}\label{sec:optimization}
We present here the optimization problem for the different controllers. Note that in all the three cases we obtain a standard quadratic programming problem.
\subsubsection{Centralized problem (CentMPC)}
The optimization problem for the whole constellation is:
\small
\begin{mini}
  {\bm{u},\bm{x}}{\summa{i=1}{N} \summa{k=1}{N\rs{p}-1} J_i^{\mathrm{trns}}(k) + \alpha\rs{end}J^{\mathrm{end}}_i(N\rs{p})}{\label{eq:optProblemRingCent}}{}
  \addConstraint{x_i(k+1)}{= Ax_i(k) + Bu_i(k),}{\quad i=1,\ldots,M}
  \addConstraint{}{}{\quad k=0,\ldots,N\rs{p}-1}
  \addConstraint{\|\bm{u}(k)\|_\infty} {\leq u\rs{max},}{\quad k=0,\ldots,N\rs{p}-1}{}
  \addConstraint{\bm{x}}{= \bm{x_0},}{}{}{}
\end{mini}
\normalsize
where $x_0$ are the initial states at time step $0$, $\bm{u} = \begin{bmatrix} \bm{u}_1\tr \dots \bm{u}_N\tr\end{bmatrix}\tr$, with $\bm{u}_i = \begin{bmatrix} u_i(0) \dots u_i(N\rs{p-1})\end{bmatrix}$ ; $\bm{x}$ and $\bm{x}_0$ are defined similarly. 

\subsubsection{Fully decentralized problem (FD-MPC)}
The optimization problem for each single spacecraft is:
\begin{mini}[2]
{\bm{u}_i,\bm{x}_i}{\summa{k=1}{N\rs{p}-1} J_i^{\mathrm{trns}}(k) + \alpha\rs{end}J^{\mathrm{end}}_i(N\rs{p})}
{\label{eq:optProblemRingFullDec}}{}
    \addConstraint{x_i(k+1)}{= Ax_i(k) + Bu_i(k),}{\quad k=0,\ldots,N\rs{p}-1}
  \addConstraint{\|\bm{u}(k)\|_\infty} {\leq u\rs{max},}{\quad k=0,\ldots,N\rs{p}-1}{}
  \addConstraint{\bm{x}(0)}{= \bm{x}_{0},}{}{}{}
\end{mini}

Note that, w.r.t.\ \eqref{eq:optProblemRingCent}, in \eqref{eq:optProblemRingFullDec} only the terms related to each single spacecraft are present.
In the fully decentralized case, only the initial state of the neighbors is known. Therefore, in \eqref{eq:thetaSetpoint}, $\theta_j(k) = \theta_j(0)$, $\forall j \in \mathcal{N}_i$, $\forall k = 0,\dots,N\rs{p}-1$.

\subsubsection{Decentralized problem with information sharing (DMPC-IS)}
The optimization problem is the same as \eqref{eq:optProblemRingFullDec}, with one main difference. In this case, the spacecraft can communicate and thus their full trajectory is shared with their neighbors. Therefore, in \eqref{eq:thetaSetpoint}, $\theta_j = \theta_j^*$, $\forall j \in \mathcal{N}_i$, $\forall i = 1,\dots,N$, where $\theta_j^*$ is the optimal trajectory of $\theta_j$ obtained at the previous time step. For the initial time step, when no other optimal trajectory is available, $\theta_j(0)$ can be used instead.

\section{Numerical simulations} \label{sec:simulations}

\subsection{Scenario}
We consider a reconfiguration scenario of spacecraft in a circular formation. The $N\rs{sat}$ satellites are equally spaced and at the beginning of the simulation $N\rs{deorbit}$ satellites leave the orbit. The remaining ones have to reconfigure their position so that an equidistant formation is achieved again.  

\subsection{Setup}
With this scenario we want to investigate how the performance of the designed controllers of Section \ref{sec:control}, i.e., centralized MPC (CentMPC), fully decentralized MPC (FD-MPC), and decentralized MPC with information sharing (DMPC-IS), compare to each other. Moreover, we study the change in performance as the number of neighbors in \eqref{eq:thetaSetpoint} considered by each spacecraft increases. To do so, at each simulation we obtain an initial configuration for the constellation and we perform a full simulation for each controller by changing only the amount of neighbors. We perform then several simulations and average the results.

The mass of the spacecraft is $m$, their maximum thrust is $u\rs{max}$ and the chosen orbit has an orbital altitude of $R\rs{orbit}$ and and orbital period of $T\rs{orbit}$. 
The list of all the parameters is shown in Table \ref{tab:ringSimulationParameters}.

All the simulations were performed on MATLAB version R2020b, on a computer with 32 GB of RAM and an Intel  i5-6500 processor. The solver used is CPLEX \cite{cplex2009}, but any other quadratic programming solver could have been used, e.g.\ OSQP \cite{osqp}. We also used YALMIP \cite{lofberg2004} to model the optimization problem.

\begin{table}[t]
\begin{center}
\begin{tabular}{c|c|c} 
    Symbol & Meaning & Value\\
 \hline
$T\rs{s}$ & Sampling time & $355$ (s) \\
$N\rs{p}$ & Prediction horizon & $64$ (4 orbits)\\
$N\rs{sat}$ & \# of spacecraft & $40$\\
$N\rs{deorbit}$ & \# of deorbited spacecraft  & $5$\\
$\alpha\rs{end}$ & Final position error weight & $10^5$\\
$\left[\alpha_u,\alpha_\theta, \alpha_\rho \right]$ & Cost function weights  & $\left[2.5\cdot 10^{-3}, 1, 10^{-5}\right]$\\
$N\rs{loops}$ & Number of simulation steps & $400$ (25 orbits)  \\
$u\rs{max}$ & Maximum thrust & $200$ (mN)  \\
$m$ & Spacecraft mass & $200$ (kg)  \\
$R\rs{orbit}$ & Orbit altitude & $500$ (km) \\
 \end{tabular}
\end{center}
\caption{Simulation parameters}  
\label{tab:ringSimulationParameters}
\end{table}

\subsection{Closed-loop simulation model}
At each simulation time step, we solve an optimal control problem, i.e., either \eqref{eq:optProblemRingCent} or \eqref{eq:optProblemRingFullDec}, according to the controller used. We then obtain an optimal input sequence for the actuators of the spacecraft. At this point, in order simulate the evolution of the real physical system, we apply the optimal inputs to a more realistic model than the one used for the optimization problem. Therefore, to describe the motion of the spacecraft around the Earth, we use the model
\begin{equation}\label{eq:cartmotion}
    \bm{\ddot r} =  -\mu \frac{\bm{r}}{r^3} + \bm{a_F}
\end{equation}
where $r$ is the magnitude of the position vector $\bm{r}$ of the spacecraft in the ECI frame, $\mu$ is the gravitational parameter of the Earth, $\bm{\ddot r}$ is the acceleration of the position vector, and $\bm{a_F}$ comprises all accelerations due to other forces, including control force. 
This model allows us to rapidly simulate the motion of the spacecraft in space. Then, we propagate the states of the spacecraft for one time step using the optimal inputs found, employing the ODE45 function of MATLAB, which yields the values of the next states of the spacecraft. These states are provided to the new optimization problem as initial states of the spacecraft. 


\subsection{Performance indices}\label{sec:performanceIndices}
We use two different performance indices to evaluate the performance of the controllers and the impact of some parameters.
The first performance index is the average of the absolute difference of the distance between a spacecraft and its two closest neighbors at the end of the simulation period and the actual desired distance from the two closest neighbors. In other words, we  check at the values, at the end of the simulation, of $D_i = \frac{1}{2}\left( d(x_i,x_{i+1}) + d(x_i,x_{i-1})\right)$, where $d(x_i,x_{i+1})$ is the distance between spacecraft $i$ and $i+1$. Spacecraft indexed as $i-1$ and $i+1$ are the two closest neighbors of spacecraft $i$. Then, we compute the distance $\bar{D}$ that any two spacecraft would have in an equidistant formation. Finally, we compute the mean of the absolute positioning error at the end of the simulation, i.e.\ $\mu\rs{sim} = \summa{i=1}{N} |D_i-\bar{D}|$. We then average this value for all simulations, i.e.\ $\bm{\mu} = \summa{j=1}{\text{\# sim}} \mu\rs{sim}$. For a single simulation, consider simply $\mu\rs{sim}$. We also compute the maximum absolute positioning error in a similar way. If this quantity is small, it means that the positioning error at the end of the simulation period is small and thus the spacecraft have managed to reconfigure to an equidistant formation successfully.

The second performance index is the total absolute value of the input used in the constellation. This value should be as small as possible to limit the fuel consumption. We compute both the average value and the maximum for all the simulations.

\subsection{Results and discussion}

\begin{figure}[t]
    \centering
    \includegraphics[width=0.9\linewidth]{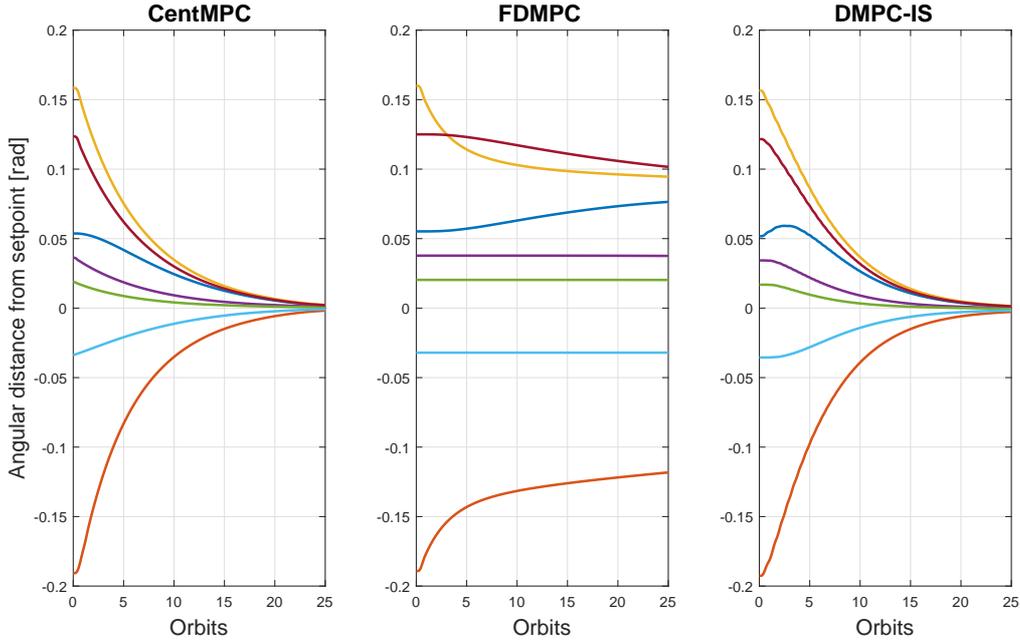}
    \caption{Angular difference from setpoint per orbit.}
    \label{fig:angularDifference}
\end{figure}

\begin{figure}[t]
    \centering
    \includegraphics[width=0.9\linewidth]{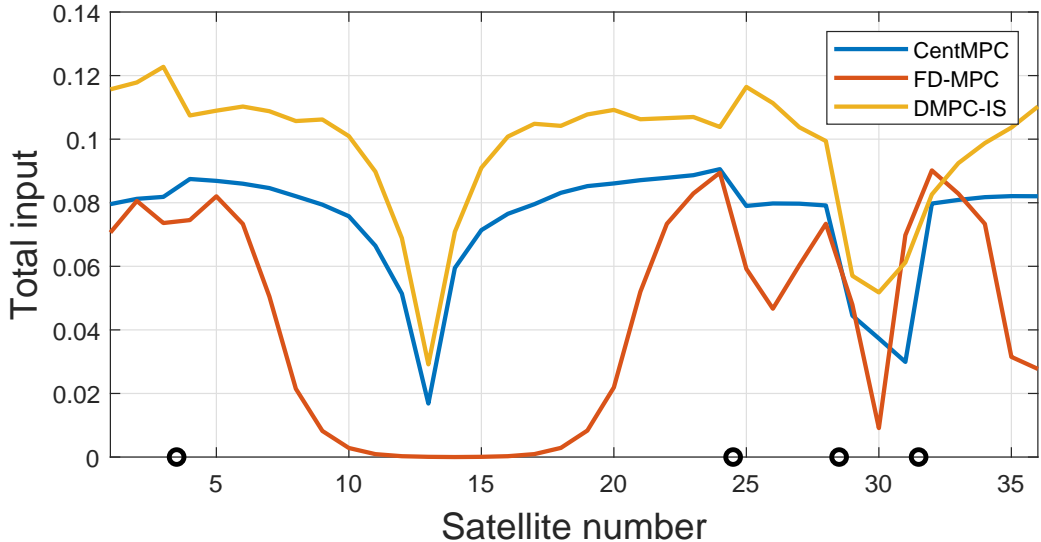}
    \caption{Total input per satellite. The black circles on the $x$-axis indicate the positions of spacecraft that have been deorbited.}
    \label{fig:totalInputRing}
\end{figure}

\subsubsection{Single simulation}
We present here a single simulation using the parameters of Table \ref{tab:ringSimulationParameters}. We analyze the results in Figures \ref{fig:angularDifference}--\ref{fig:totalInputRing}.
In Figure \ref{fig:angularDifference}, we show the angular distance of six out of 35 spacecraft from their desired relative setpoint, computed as explained in Section \ref{sec:performanceIndices}, for the three controllers. Only six spacecraft are shown to avoid confusion. Nevertheless, we can notice some interesting convergence characteristics. As expected, CentMPC is the fastest to converge to the final position, followed by the DMPC-IS and lastly by the FD-MPC. We can notice the large delay in the FD-MPC: since only the initial state is known, the satellites need a large number of orbits to converge. On the other hand, the DMPC-IS converges in a much faster way, almost in the same amount of time required by the CentMPC.
However, we can notice some oscillations in the convergence of both FD-MPC and DMPC-IS. In particular, we can look at the spacecraft in blue color in Figure \ref{fig:angularDifference}, and notice how before converging, the spacecraft first increases its angular distance from the setpoint and then it converges. One way to mitigate this is to increase the number of neighbors $\mathcal{N}$, so that more information is shared across the constellation and therefore spacecraft would more precisely anticipate how to reconfigure, avoiding a more ``reactive'' kind of control action.

The total input consumption per spacecraft is shown in Figure \ref{fig:totalInputRing}. We can clearly see how the FD-MPC is the one that consumes less fuel compared with the other two strategies, but this is only because the satellites do not move too much during the simulation, as seen in Figure \ref{fig:angularDifference}. On the other hand, DMPC-IS consumes more fuel than the CentMPC one, while having a similar convergence characteristic. This is as expected, since the CentMPC has more information available than the DMPC-IS. However, it is clear how sharing the whole nominal trajectory allows the DMPC-IS controller to achieve a much better performance than the FD-MPC controller. Lastly, the black circles in the $x$-axis indicate the positions of the spacecraft that are deorbited. We can appreciate how the spacecraft that are close to a satellite that has deorbited consume more fuel since they need to close the gap in the formation, unless the spacecraft are between two gaps, as it happens to satellite 30, which consumes less fuel than its neighbors.

\subsubsection{Comparison with different number of neighbors}
\begin{figure}[t]
    \centering
    \includegraphics[width=0.9\linewidth]{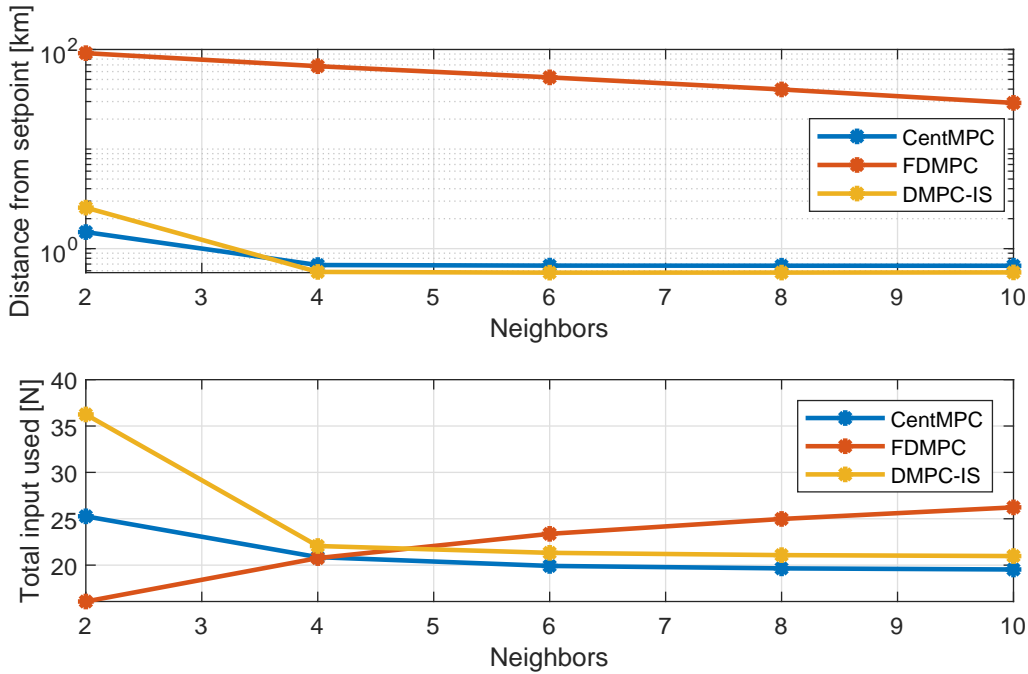}
    \caption{Relative position error averaged for all the simulations in the top plot, with the y-axis in log scale, and total input averaged for all the simulations in the bottom plot.}
    \label{fig:positionErrorTotalInputLog}
\end{figure}

\begin{table}[t]
\begin{center}
\begin{tabular}{c|c|c|c} 
   $|\mathcal{N}|$ & CentMPC & FD-MPC & DMPC-IS\\
\hline

2 & \begin{tabular}{c}
$\mu = 1.466$  \\ $\sigma = 0.942$
\end{tabular}  & \begin{tabular}{c}
$\mu = 91.878$  \\ $\sigma = 32.648$
\end{tabular}  & \begin{tabular}{c}
$\mu = 2.582$  \\ $\sigma = 2.128$
\end{tabular} \\

4 & \begin{tabular}{c}
$\mu = 0.684$  \\ $\sigma = 0.028$
\end{tabular}  & \begin{tabular}{c}
$\mu = 67.957$  \\ $\sigma = 29.708$
\end{tabular}  & \begin{tabular}{c}
$\mu = 0.584$  \\ $\sigma = 0.046$
\end{tabular} \\

6 & \begin{tabular}{c}
$\mu = 0.675$  \\ $\sigma = 0.029$
\end{tabular}  & \begin{tabular}{c}
$\mu = 52.476$  \\ $\sigma = 24.136$
\end{tabular}  & \begin{tabular}{c}
$\mu = 0.574$  \\ $\sigma = 0.031$
\end{tabular} \\

8 & \begin{tabular}{c}
$\mu = 0.674$  \\ $\sigma = 0.033$
\end{tabular}  & \begin{tabular}{c}
$\mu = 39.680$  \\ $\sigma = 18.735$
\end{tabular}  & \begin{tabular}{c}
$\mu = 0.575$  \\ $\sigma = 0.029$
\end{tabular} \\

10 & \begin{tabular}{c}
$\mu = 0.673$  \\ $\sigma = 0.034$
\end{tabular}  & \begin{tabular}{c}
$\mu = 29.082$  \\ $\sigma = 13.812$
\end{tabular}  & \begin{tabular}{c}
$\mu = 0.577$  \\ $\sigma = 0.028$
\end{tabular} \\
\end{tabular}
\end{center}
\caption{Simulations results - Position error at the end of the simulation, measured in [km]}  
\label{tab:ringScenarioNeighborsPosition}
\end{table}

\begin{table}[t]
\begin{center}
\begin{tabular}{c|c|c|c} 
   $|\mathcal{N}|$ & CentMPC & FD-MPC & DMPC-IS\\
\hline

2 & \begin{tabular}{c}
$\mu = 2.526$  \\ $\sigma = 0.258$
\end{tabular}  & \begin{tabular}{c}
$\mu = 1.608$  \\ $\sigma = 0.318$
\end{tabular}  & \begin{tabular}{c}
$\mu = 3.625$  \\ $\sigma = 0.836$
\end{tabular} \\

4 & \begin{tabular}{c}
$\mu = 2.086$  \\ $\sigma = 0.236$
\end{tabular}  & \begin{tabular}{c}
$\mu = 2.076$  \\ $\sigma = 0.300$
\end{tabular}  & \begin{tabular}{c}
$\mu = 2.206$  \\ $\sigma = 0.191$
\end{tabular} \\

6 & \begin{tabular}{c}
$\mu = 1.991$  \\ $\sigma = 0.207$
\end{tabular}  & \begin{tabular}{c}
$\mu = 2.337$  \\ $\sigma = 0.222$
\end{tabular}  & \begin{tabular}{c}
$\mu = 2.133$  \\ $\sigma = 0.188$
\end{tabular} \\

8 & \begin{tabular}{c}
$\mu = 1.965$  \\ $\sigma = 0.196$
\end{tabular}  & \begin{tabular}{c}
$\mu = 2.498$  \\ $\sigma = 0.139$
\end{tabular}  & \begin{tabular}{c}
$\mu = 2.108$  \\ $\sigma = 0.187$
\end{tabular} \\

10 & \begin{tabular}{c}
$\mu = 1.953$  \\ $\sigma = 0.203$
\end{tabular}  & \begin{tabular}{c}
$\mu = 2.623$  \\ $\sigma = 0.113$
\end{tabular}  & \begin{tabular}{c}
$\mu = 2.098$  \\ $\sigma = 0.187$
\end{tabular} \\
\end{tabular}
\end{center}
\caption{Simulations results - Total input}  
\label{tab:ringScenarioNeighborsInput}
\end{table} 

\begin{table}[t]
\begin{center}
\begin{tabular}{c|c} 
    Controller &  Total Time [s] \\
 \hline
CentMPC & $3422.7$\\
 
FD-MPC & $342.2$\\
 
DMPC-IS & $346.3$\\
 \end{tabular}
\end{center}
\caption{Simulations results - Total optimization time}  
\label{tab:ringScenarioNeighborsTime}
\end{table}

We study here the effect of a different number of neighbors, $\mathcal{N}$, on the performance of the three chosen controllers. We consider 10 different simulation scenarios, in each of which a different configuration of spacecraft leaves the orbit, and for each of them we run 5 sub-simulations where everything is kept the same except for $\mathcal{N}$. This is done in order to properly analyze the effect of $\mathcal{N}$ on the performance of the controllers. If we did not keep the initial conditions the same while changing $\mathcal{N}$ at each sub-simulation, we would reach wrong conclusions on the real performance change. Once all the sub-simulations are finished, we go to the next simulation, were we draw new initial conditions.

We evaluate the two different metrics mentioned in Section \ref{sec:performanceIndices}. The results are shown in Tables \ref{tab:ringScenarioNeighborsPosition} and \ref{tab:ringScenarioNeighborsInput}, for the positioning error and the total input used by the constellation, respectively. We indicate the mean value by $\mu$ and the standard deviation by $\sigma$. Moreover, we show in Table \ref{tab:ringScenarioNeighborsTime} the average of the total computation time per simulation. From the results, we can see that increasing the number of neighbors improves the performance considerably. In Table \ref{tab:ringScenarioNeighborsPosition} we can see that the position error after 25 orbits becomes always smaller as the number of neighbors increases. The behavior is not always monotonic, e.g., for the DMPC-IS case $|\mathcal{N}|=10$ provides a slightly worse performance than for $|\mathcal{N}|=6$ and $|\mathcal{N}|=8$, although the difference is minor. However, the overall trend is clear, i.e., the error decreases as more neighbors are considered. Table \ref{tab:ringScenarioNeighborsInput} provides a more clear picture: the total input used keeps decreasing in a quite noticeable way as we increase the number of neighbors, for the CentMPC and DMPC-IS strategies. It might not be intuitive why the performance improves for the CentMPC case, as such controller already has information on the whole constellation. However, note that the cost defined in \eqref{eq:optProblemRingCent} depends on the amount of neighbors considered, so that a better setpoint can be computed when more neighbors are taken into account. 
For the FD-MPC case instead, the total input increases, but this extra input used allows the controller to achieve a better positioning error. The results of Tables \ref{tab:ringScenarioNeighborsPosition}, \ref{tab:ringScenarioNeighborsInput} are also depicted in Figure \ref{fig:positionErrorTotalInputLog} (respectively in the top and bottom plot), where in the top plot the y-axis is in log-scale. We can notice once again how increasing the number of neighbors leads to a reduction of both the position error and the total input.

We can thus claim that adding more neighbors, i.e., sharing more information among the spacecraft, improves the convergence speed to the equidistant formation and reduces the fuel consumption. Having more information of where the neighbors currently are (FD-MPC case) or are going to be (DMPC-IS case) leads to less oscillations of the spacecraft and it makes them point directly to their final setpoint. Moreover, as the number of neighbors increases, the solution provided by DMPC-IS becomes closer and closer to the one of CentMPC. It is also interesting to confirm, as expected, that the FD-MPC yields always worse performances than DMPC-IS. This confirms the importance of having some sort of inter-satellite communication in the constellation, rather than simple ``sensing''.

Lastly, the total computation time shown in Table \ref{tab:ringScenarioNeighborsTime}, shows how CentMPC requires around 10 times the computation time required by the other two strategies. Increasing the number of spacecraft in the constellation will keep increasing this gap, which confirms that centralized strategies are not suitable for online computation in spacecraft large constellations and non-centralized solutions are needed.

\section{Conclusions and Future Work} \label{sec:conclusions}
We presented a comparison between a centralized controller and two decentralized ones for a reconfiguration problem of a spacecraft constellation in circular orbit formation. We suppose that some of the spacecraft leave the orbit and the other ones have to reconfigure so as to achieve an equidistant formation. The two decentralized strategies differ in the amount of information available to every spacecraft, namely either the current state or the whole planned trajectory of their neighbors. The simulation results show the benefits of an increased amount of information shared, i.e., as information from more neighbors is available, the spacecraft manage to converge in a faster way to the desired formation, while using also less fuel. Moreover, both decentralized strategies outperform the centralized one in terms of the computation time. For large formations, computing a centralized control action in real time might be infeasible.

Future research will focus on distributed strategies, in which every spacecraft solves a joint optimization problem in cooperation with its neighbors. Moreover, we will perform a similar study related to a attitude coordination problem. Lastly, we will consider a high-fidelity model including several disturbances for the closed-loop simulation model. 

\bibliography{references}

\begin{thebibliography}{10}

\bibitem{starlink2020}
{SpaceX - Starlink}.
\newblock \url{https://www.starlink.com}, 2020.

\bibitem{iridium2021}
{Iridium}.
\newblock \url{https://www.iridium.com}, 2021.

\bibitem{vandam2019b}
F.~van Dam.
\newblock {Distributed collision free trajectory optimization for the
  reconfiguration of a spacecraft formation}.
\newblock Master's thesis, Delft University of Technology, the Netherlands,
  2019.

\bibitem{hadaegh2016}
F.~Y. Hadaegh, S.-J. Chung, and H.~M. Manohara.
\newblock {On development of 100-gram-class spacecraft for swarm applications}.
\newblock {\em IEEE Systems Journal}, 10(2):673--684, 2016.
\newblock
  \href{https://doi.org/10.1109/JSYST.2014.2327972}{\color{blue}{\underline{DOI:\,10.1109/JSYST.2014.2327972}}}.

\bibitem{diMauro2018}
G.~{Di Mauro}, M.~Lawn, and R.~Bevilacqua.
\newblock {Survey on guidance navigation and control requirements for
  spacecraft formation-flying missions}.
\newblock {\em Journal of Guidance, Control, and Dynamics}, 41(3):581--602,
  2018.
\newblock
  \href{https://doi.org/10.2514/1.G002868}{\color{blue}{\underline{DOI:\,10.2514/1.G002868}}}.

\bibitem{bandyopadhyay2016b}
S.~Bandyopadhyay, R.~Foust, G.~P. Subramanian, S.-J. Chung, and F.~Y. Hadaegh.
\newblock {Review of formation flying and constellation missions using
  nanosatellites}.
\newblock {\em Journal of Spacecraft and Rockets}, 53(3):567--578, 2016.
\newblock
  \href{https://doi.org/10.2514/1.A33291}{\color{blue}{\underline{DOI:\,10.2514/1.A33291}}}.

\bibitem{sinYin2020}
E.~Sin, H.~Yin, and M.~Arcak.
\newblock Passivity-based distributed acquisition and station-keeping control
  of a satellite constellation in areostationary orbit.
\newblock In {\em Dynamic Systems and Control Conference}, volume 84287.
  American Society of Mechanical Engineers, 2020.
\newblock
  \href{https://doi.org/10.1115/DSCC2020-3136}{\color{blue}{\underline{DOI:\,10.1115/DSCC2020-3136}}}.

\bibitem{vos2014}
E.~Vos, J.~M.~A. Scherpen, and A.~J. van~der Schaft.
\newblock Equal distribution of satellite constellations on circular target
  orbits.
\newblock {\em Automatica}, 50(10):2641--2647, 2014.
\newblock
  \href{https://doi.org/10.1016/j.automatica.2014.08.027}{\color{blue}{\underline{DOI:\,10.1016/j.automatica.2014.08.027}}}.

\bibitem{morgan2014a}
D.~Morgan, S.-J. Chung, and F.~Y. Hadaegh.
\newblock {Model predictive control of swarms of spacecraft using sequential
  convex programming}.
\newblock {\em Journal of Guidance, Control, and Dynamics}, 37(6):1725--1740,
  2014.
\newblock
  \href{https://doi.org/10.2514/1.G000218}{\color{blue}{\underline{DOI:\,10.2514/1.G000218}}}.

\bibitem{morgan2016}
D.~Morgan, G.~P. Subramanian, S.-J. Chung, and F.~Y. Hadaegh.
\newblock {Swarm assignment and trajectory optimization using variable-swarm,
  distributed auction assignment and sequential convex programming}.
\newblock {\em International Journal of Robotics Research}, 35(10):1261--1285,
  2016.
\newblock
  \href{https://doi.org/10.1177/0278364916632065}{\color{blue}{\underline{DOI:\,10.1177/0278364916632065}}}.

\bibitem{foust2020a}
R.~Foust, E.~Lupu, Y.~Nakka, S.-J. Chung, and Hadaegh F.
\newblock {Autonomous in-orbit satellite assembly from a modular heterogeneous
  swarm}.
\newblock {\em Acta Astronautica}, 169(May 2019):191--205, 2020.
\newblock
  \href{https://doi.org/10.1016/j.actaastro.2020.01.006}{\color{blue}{\underline{DOI:\,10.1016/j.actaastro.2020.01.006}}}.

\bibitem{chen2018}
H.~Chen, G.~Ning, J.~Sun, K.~Li, H.~Liu, and S.~Zhang.
\newblock Distributed and coordinated control for large scale spacecraft swarm
  using sliding mode control and artificial bifurcating potential field.
\newblock In {\em 2018 AIAA Guidance, Navigation, and Control Conference}, page
  1861, 2018.
\newblock
  \href{https://doi.org/10.2514/6.2018-1861}{\color{blue}{\underline{DOI:\,10.2514/6.2018-1861}}}.

\bibitem{esaavoidance}
{ESA spacecraft dodges large constellation}.
\newblock
  \url{https://www.esa.int/Safety_Security/ESA_spacecraft_dodges_large_constellation},
  2019.

\bibitem{montenbruck2011}
O.~Montenbruck, M.~Wermuth, and R.~Kahle.
\newblock {GPS} based relative navigation for the {TanDEM-X} mission-first
  flight results.
\newblock {\em Navigation}, 58(4):293--304, 2011.
\newblock
  \href{https://doi.org/10.1002/j.2161-4296.2011.tb02587.x}{\color{blue}{\underline{DOI:\,10.1002/j.2161-4296.2011.tb02587.x}}}.

\bibitem{negenborn2014}
R.~R. {Negenborn} and J.~M. {Maestre}.
\newblock {Distributed Model Predictive Control: An Overview and Roadmap of
  Future Research Opportunities}.
\newblock {\em IEEE Control Systems Magazine}, 34(4):87--97, 2014.
\newblock
  \href{https://doi.org/10.1109/MCS.2014.2320397}{\color{blue}{\underline{DOI:\,10.1109/MCS.2014.2320397}}}.

\bibitem{morgan2012b}
D.~Morgan, S.-J. Chung, and F.~Y. Hadaegh.
\newblock Spacecraft swarm guidance using a sequence of decentralized convex
  optimizations.
\newblock In {\em AIAA/AAS astrodynamics specialist conference}, page 4583,
  2012.
\newblock
  \href{https://doi.org/10.2514/6.2012-4583}{\color{blue}{\underline{DOI:\,10.2514/6.2012-4583}}}.

\bibitem{koenig2018}
A.~W. Koenig and S.~D'Amico.
\newblock {Robust and safe N-spacecraft swarming in perturbed near-circular
  orbits}.
\newblock {\em Journal of Guidance, Control, and Dynamics}, 41(8):1643--1662,
  2018.
\newblock
  \href{https://doi.org/10.2514/1.G003249}{\color{blue}{\underline{DOI:\,10.2514/1.G003249}}}.

\bibitem{camacho2013}
E.~F. Camacho and C.~B. Alba.
\newblock {\em Model Predictive Control}.
\newblock Advanced Textbooks in Control and Signal Processing. Springer London,
  2013.
\newblock
  \href{https://doi.org/10.1007/978-0-85729-398-5}{\color{blue}{\underline{DOI:\,10.1007/978-0-85729-398-5}}}.

\bibitem{debruijn2011}
F.~de~Bruijn, E.~Gill, and J.~How.
\newblock Comparative analysis of cartesian and curvilinear clohessy-wiltshire
  equations.
\newblock {\em Journal of Aerospace Engineering}, 3(2):1, 2011.

\bibitem{geller2017}
D.~K. Geller and T.~A. Lovell.
\newblock Angles-only initial relative orbit determination performance analysis
  using cylindrical coordinates.
\newblock {\em The Journal of the Astronautical Sciences}, 64(1):72--96, 2017.
\newblock
  \href{https://doi.org/10.1007/s40295-016-0095-z}{\color{blue}{\underline{DOI:\,10.1007/s40295-016-0095-z}}}.

\bibitem{hartley2015}
E.~N. Hartley.
\newblock A tutorial on model predictive control for spacecraft rendezvous.
\newblock In {\em 2015 European Control Conference (ECC)}, pages 1355--1361.
  IEEE, 2015.
\newblock
  \href{https://doi.org/10.1109/ECC.2015.7330727}{\color{blue}{\underline{DOI:\,10.1109/ECC.2015.7330727}}}.

\bibitem{cplex2009}
IBM~ILOG Cplex.
\newblock V12. 1: User’s manual for {CPLEX}.
\newblock {\em International Business Machines Corporation}, 46(53):157, 2009.

\bibitem{osqp}
B.~Stellato, G.~Banjac, P.~Goulart, A.~Bemporad, and S.~Boyd.
\newblock {OSQP}: an operator splitting solver for quadratic programs.
\newblock {\em Mathematical Programming Computation}, 12(4):637--672, 2020.
\newblock
  \href{https://doi.org/10.1007/s12532-020-00179-2}{\color{blue}{\underline{DOI:\,10.1007/s12532-020-00179-2}}}.

\bibitem{lofberg2004}
J.~L{\"{o}}fberg.
\newblock {YALMIP} : A toolbox for modeling and optimization in matlab.
\newblock In {\em In Proceedings of the CACSD Conference}, Taipei, Taiwan,
  2004.
\newblock
  \href{https://doi.org/10.1109/CACSD.2004.1393890}{\color{blue}{\underline{DOI:\,10.1109/CACSD.2004.1393890}}}.

\end{thebibliography}

\end{document}